\documentstyle[psfig,sprocl]{article}
\newcommand{\be}{\begin{equation}}
\newcommand{\ee}{\end{equation}}
\newcommand{\bea}{\begin{eqnarray}}
\newcommand{\eea}{\end{eqnarray}}
\newcommand{\bean}{\begin{eqnarray*}}
\newcommand{\eean}{\end{eqnarray*}}
\newcommand{\non}{\nonumber}
\newcommand{\ra}{\rightarrow}

\newcommand{\bc}{\begin{center}}
\newcommand{\ec}{\end{center}}

\newcommand{\ba}{\begin{array}}
\newcommand{\ea}{\end{array}}

\begin{document}  
\vspace*{0.5cm}
                                                            
\title{ELECTROWEAK SUDAKOV CORRECTIONS\\ AT 2 LOOP LEVEL
\footnote{Talk presented at PPP2000                         
workshop at Chipen, Taiwan, November 2000 }}               
                                                            
\author{HIROYUKI KAWAMURA\footnote{JSPS Research Fellow}}   
                                                            
\address{ Department of Physics, Hiroshima University\\     
          Higashi-Hiroshima 739-8526, JAPAN\\               
          E-mail:kawamura@theo.phys.sci.hiroshima-u.ac.jp}    
                                                              
\maketitle

\begin{abstract}             
{
\normalsize
\noindent                 
In processes at the energy much higher than electroweak scale, 
weak boson mass act as the infrared cutoff in weak boson loops 
and resulting Sudakov log corrections can be as large as 10\%. 
Since electroweak theory is off-diagonally broken gauge theory, 
its IR structure is quite different from that of QCD.
We briefly review recent developments on electroweak Sudakov and 
discuss on the exponentiation of Sudakov double logs and 
explicit 2 loop calculations in Feynman gauge.
}
\end{abstract}


\section{Introduction}
High energy experiments in TeV region are planned in the near future 
to obtain the hints to the fundamental problems       
in particle physics such as the mechanism of electroweak symmetry  
breaking, the gauge hierarchy problem and so on.  
The precision measurements in this region are expected to give us   
the important information to construct the scenario favorable up to   
Plank scale.\footnote{\normalsize The possibility of the large extra dimension 
is not considered in this note.} 
In order to extract these information from experimental data, 
it is crucial to carry out the theoretical calculations at least 
in the same level of accuracy in the standard model and 
in other possible models. It is very important also for the
estimation of the background of the production of the new particle. 
                     
Recently Ciafaloni and Comelli \cite{CC1,CCC} pointed out that in 
processes with energy much higher than electroweak scale, the 
``heavy particle masses'' $M_W \simeq M_Z ( \equiv M) $ acts as 
IR cutoff in weak boson loop integrals and resulting IR log 
corrections give large contributions to cross sections, comparable 
or larger than QCD corrections. For example, 1 loop log corrections 
to $e^+e^- \ra \mu^+\mu^-$ are \cite{BCCRV}, 
\bea
\sigma^\mu=\sigma^{tree}\times 
\left[1+\displaystyle{\frac{\alpha}{4\pi\sin^2\theta_W}}
\left\{0.6\ln\frac{Q^2}{\mu^2}
+9.4\ln{\frac{Q^2}{M^2}}-1.4\ln^2\frac{Q^2}{M^2}
\right\}
\right].
\non
\eea
\vspace{-0.6cm}

{\footnotesize\hspace*{4.7cm} UV log \hspace{0.5cm} IR log   
\hspace{0.6cm} Sudakov log}\\

\noindent 
The second and the third terms which vanish in LEP energy region 
give the dominant corrections in TeV region.
	
Especially Sudakov double logarisms \cite{SU} which come 
from the overlap of the soft and the collinear region of the loop integral 
give the dominant radiative corrections in the asymptotic regime	
(Their typical value at  
$\sqrt{S}\simeq 1TeV$ is $\frac{\alpha}{4\pi\sin^2{\theta}}           
\log^2{\frac{S}{M_W^2}}\simeq 6.6\% $).
Since factors like $\alpha^n\log^{2n}\frac{S}{M_W^2}$ (LL),
$\alpha^n\log^{2n-1}\frac{S}{M_W^2}$ (NLO), $\cdots$, 
may spoil the perturbation expansion, 
it is necessary to control or resum these contributions     
to obtain a reliable and precise calculation in extremely high energy 
processes.

The infrared structure of gauge theory has been extensively
investigated and it is well-known that Sudakov logs exponentiate     
for the form factor in QED \cite{SU,YFS} and QCD \cite{JCC}   
and that they can be resummed in various physical processes \cite{S}.
The exponentiation of Sudakov logs in these cases results from         
the gauge symmetry and Lorentz symmetry in the factorization    
formula of cross sections \cite{CLS}.
In case of the electroweak theory, the situation is quite different.
Since the gauge symmetry is broken and its breaking pattern          
is off-diagonal, the exponentiation of Sudakov logs in this case 
is a non-trivial matter. 
Several discussions on the all order behavior of Sudakov terms 
have been presented with different results \cite{CC2,KP,FLMM} and 
the explicit 2 loop calculations \cite{BW,M1,HKK} imply the exponentiation 
of Sudakov logs. From these discussions, we can see that  
the methods valid in QCD can not be applied straightforwardly to      
electroweak case and adequate modifications are needed  
            
The most striking feature of electroweak Sudakov is that 
they appear not only in the exclusive processes but also 
in the inclusive processes since non-abelian charge 
is not confined \cite{CCC}. 
In non-abelian gauge theory, it is known that Bloch-Nordsieck 
cancellations occur in leading power when both the sum of 
the final degenerate states and the initial color average 
are taken \cite{C}. 
In QCD prcesses where initial color averages are always taken 
due to the color confinement,  
Sudakov double logs appear only in the end-point region of the phase 
space where the soft cancellation fails due to the suppression of 
the real emissions.
On the other hand, initial state in electroweak processes 
are color (isospin) non-singlet and Sudakov logs remain even when 
the number of weak boson is not counted in the final state.  
Therefore, it can be said that the effects of symmetry breaking 
remain even we go up to the extremely high energy in which we usually 
consider that the gauge symmetry is restored.

In the following sections, we introduce 
the discussions on all order behavior of Sudakov logs and 
explain 2-loop calculation of the form factor in detail  
and discuss on other recent development.

\section{Discussion on all order behavior of Sudakov logarisms}

Discussion on all order behavior of Sudakov logarisms have been 
presented by several authors \cite{KP,CC2,FLMM}. The author of 
\cite{KP} calculated all order $\ln^{2n}{\frac{S}{M^2}}$ terms 
in the self-energy in Axial gauge including only the weak boson loop 
and showed that Sudakov terms does not exponentiate due to the  
mixing effect \cite{KP}. Ciafaloni and Comelli \cite{CC2} estimated 
the leading log terms of the form factor in Feynman gauge 
using soft insertion formula which is the formalism developed in QCD 
\cite{BCM}.
Soft insertion formula means that we can obtain leading log terms by
inserting the eikonal current at fermion legs, and estimating the 
diagrams with energy ordering. Their result was that the QED effects 
and electroweak effects factorize and Sudakov logs exponentiate 
in operator form but does not exponentiate numerically.
Fadin {\sl et.al.} \cite{FLMM} gave a general argument using 
the infrared evolution equation \cite{KL}, which is the differencial  
equation with respect to the infrared cutoff derived from Gribov's 
Bremsstrahlung theorem. They derive two equations, one is for 
the region where $\mu \leq M_W$ with QED kernel and another is 
for $\mu \geq M_W$ with electroweak kernel. 
The exponentiation as the numerical value is concluded 
naturally from their recursive form. 

\section{Explicit calculation}

Explicit 2 loop calculations of leading logarisms for the form
factor are helpful to resolve the controversy discussed above 
and to construct the rigorous all order proof of exponentiation 
of electroweak Sudakov.
We describe the similarity and the difference between this case and 
the QCD case concerning the cancellation of the non-exponential 
factors. 

\subsection{QCD case}

2 loop calculations of the leading singularities of the 
form factor were accomplished many years ago
in QED \cite{FS} and QCD \cite{FFT}.

QCD double log corrections for the form factor come from vertex 
corrections in Feynman gauge. 
Using the mass regularization for infrared divergences 
1-loop result is, \\

\hspace*{1.5cm}
\psfig{file=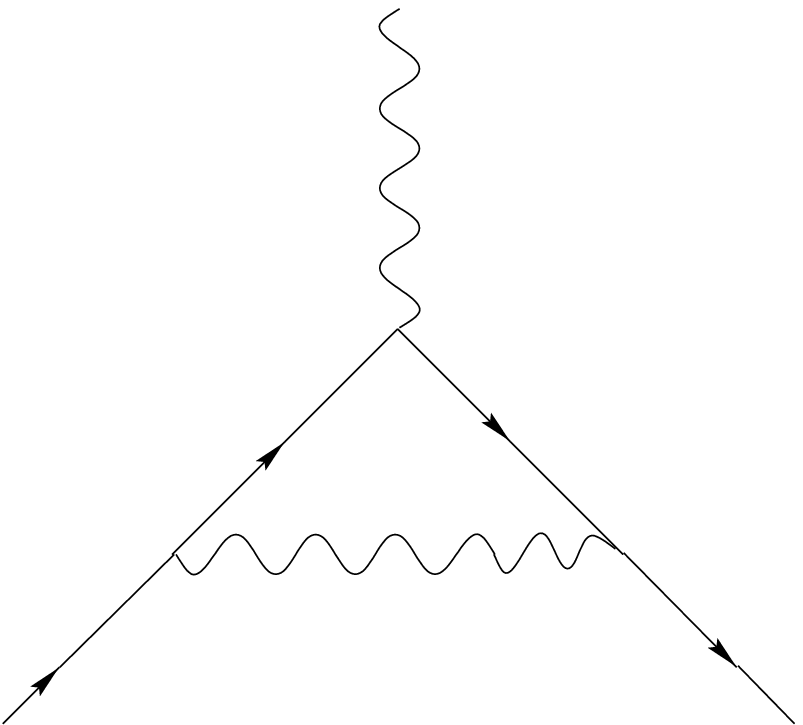,width=1.5cm}
\vspace{-1cm}

\hspace*{4cm} = \hspace{1cm}$\quad-{\displaystyle\frac{e^2Q^2}{16\pi^2}}
\ln^2{\displaystyle\frac{S}{\lambda^2}}$ \ ,\\

\vspace{0.5cm}

where $C_F$ is the SU(3) Casimir operator for the fundamental 
representation and $S$ is the momentum transfer. 
At 2 loop level, the leading double logs appear 
from the ladder diagram, the crossed ladder diagram, and the 
diagrams including triple gluon coupling as follows,\\

\hspace*{1cm}\psfig{file=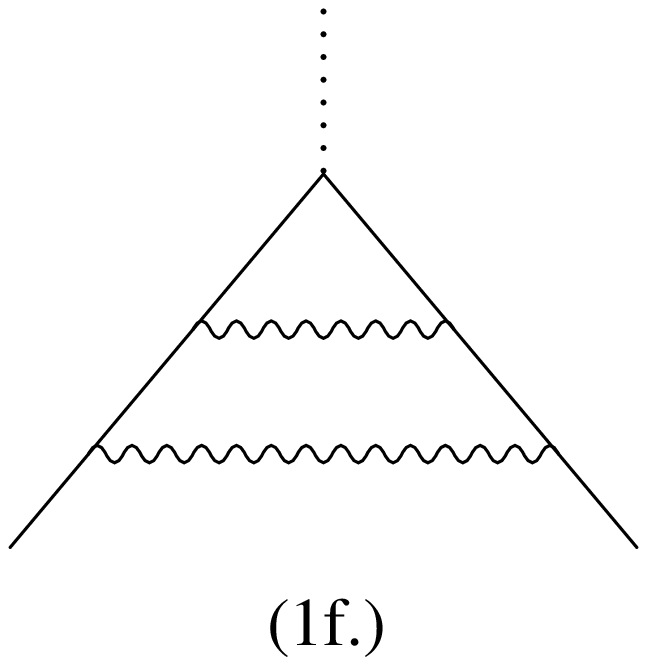,width=1.5cm}
\vspace{-1.3cm}

\hspace*{4cm} =  \hspace{0.5cm}$\quad{\displaystyle\frac{\alpha_s^2}{(4\pi)^4}}
{\displaystyle\frac{1}{24}\ln^4\frac{S}{\lambda}}
\times  C_F^2$  \ , \\
\vspace{0.5cm}

\hspace*{1cm}\psfig{file=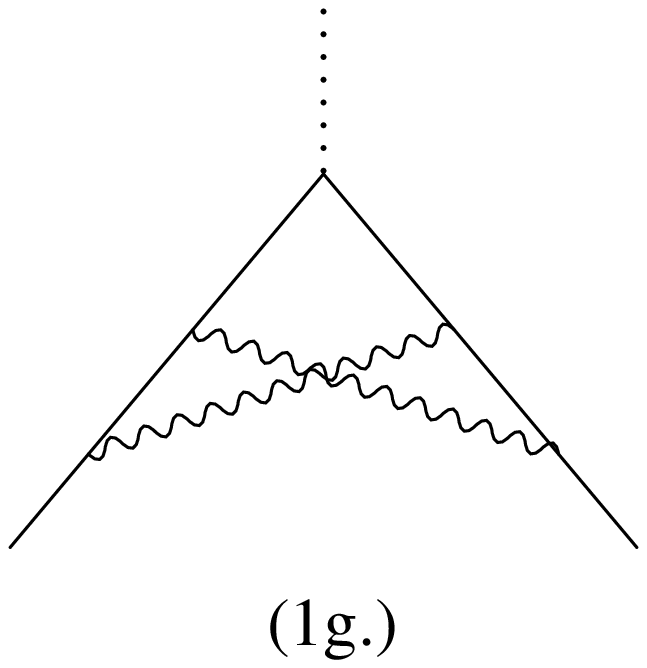,width=1.5cm}
\vspace{-1.3cm}

\hspace*{4.1cm}= \hspace{0.5cm}
$\quad {\displaystyle\frac{\alpha_s^2}{(4\pi^4)}}
{\displaystyle\frac{1}{12}}\ln^4{\displaystyle\frac{S}{\lambda^2}}
\times\left( C_F^2-\frac{1}{2}C_AC_F \right)$ \ ,\\
\vspace{0.5cm}

\hspace*{1cm}
\psfig{file=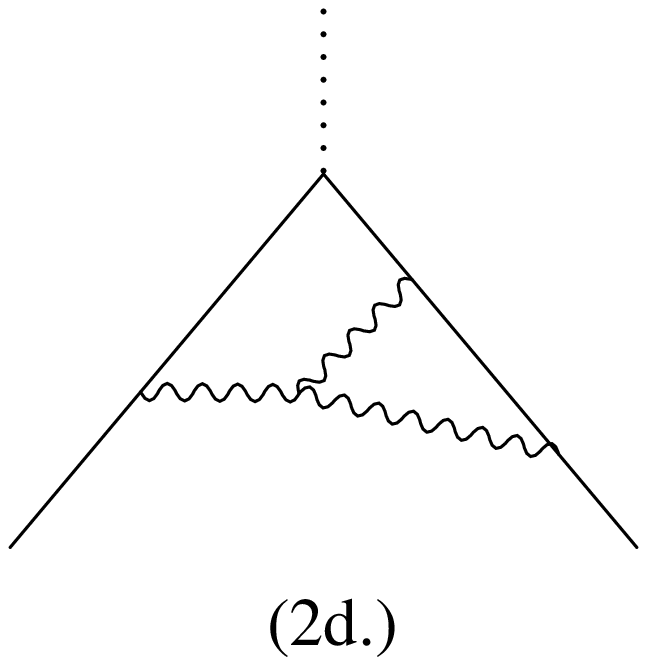,width=1.5cm} 
\vspace{-1.3cm}

\hspace*{3cm}
$\times 2$ \hspace{0.4cm} = \hspace{0.5cm}
$ -{\displaystyle\frac{\alpha_s^2}{(4\pi)^4}}
{\displaystyle\frac{1}{12}}\ln^4{\displaystyle\frac{S}{\lambda^2}}
\times\frac{1}{2}C_AC_F$ \ .                 
\vspace{1cm}

\noindent
Then the sum of these contributions are exactly a half of the 1-loop 
contribution squared.The non-exponential terms which appear 
in the crossed diagram and the diagram with the triple gluon coupling 
due to the non-abelian nature cancel each other, and the result 
is consistent with the exponentiation of the Sudakov logarisms.

\subsection{Electroweak theory}              
              
2-loop expricit calculations has been done for self energy in axial gauge 
by Beenakker and Werthenbach and for the form factor of right-handed 
fermion in Feynman gauge by Melles\cite{M1}. Here we concentrate on  
the calculations of the form factor for left-handed fermions \cite{HKK}.

The situation become more complicated in electroweak theory than QCD  
since the gauge symmetry is spontaneously broken and the pattern of 
the symmetry breaking is not diagonal. 
In physical processes photon must be treated 
in semi-inclusive way since it is uncountable, on the other hand, 
weak bosons can be treated both inclusive and exclusive ways.
The process we consider now is the fermion pair production from 
the SU(2)$\otimes$U(1) singlet source and the infrared divergences
by photon loops are regularzed by the fictitious photon mass $\lambda$.  
We must treat gauge bosons with different mass (``mass gap'').
Within the leading log approximations, the fermion chirality is conserved 
and W and Z boson mass can be approximated to be equal and the field 
of unbroken phase {\it i.e.}, W and Z can be used though 
$S \gg  M_W^M \equiv M$. 
The mixing effect comes from   
\bea
[ T^\pm , Q ] \neq 0,
\non
\eea
where $T^\pm$ is the SU(2) generator and $Q = T^3 + Y $ is 
the charge of fermion. 
For the right-handed fermion, the gauge group of electroweak 
interaction is reduced to $U(1)_Q\otimes U(1)_{Y-Q}$ and 
the exponentiation become a trivial matter.
It also should be noted that if we take $M^2 = \lambda^2$,  
the mixing effects disappear and the calculation becomes same 
with that for unbroken $SU(2)\otimes U(1)$ 
gauge theory with regularization mass $M$ and the exponentiation 
holds as QCD case. \\

The group factors for each SM boson exchange become,
\bea
\hspace*{-0.5cm}
\left\{
\begin{array}{l}
    \gamma\ {\rm exchange}\ : e^2 Q^2 \\
     W \ {\rm exchange}\ : g^2 \sum_{a= 1,2} T^a T^a \\
     Z \ {\rm exchange}\ : {\displaystyle \frac{g^2}{\cos^2 \theta_W}}
            ( T^3 - \sin^2 \theta_W Q )( T^3 - \sin^2 \theta_W Q ) \\
\hspace*{4cm} = g^2 T^3 T^3 + g'^2 Y^2 - e^2 Q^2
\end{array}
\right.
\non
\eea
Then, 1 loop result turn out to be,
\bea
\Gamma^{(1)} = 1 - \frac{1}{(4\pi)^2}
\left( g^2 C_F + g\prime^2Y^2-e^2Q^2 \right) \ln^2{\frac{S}{M^2}}
-\displaystyle{\frac{1}{(4\pi)^2}}e^2Q^2\ln^2{\frac{S}{\lambda}}.
\non
\eea
Here $C_F$ denotes the SU(2) Casimir operator for the fundamental 
representation. 
At the 2 loop level, there appear 16 diagrams which are 
classified in 3 groups. That is, ladder diagrams, crossed ladder diagrams 
(Fig.1) and the diagrams including 3 gauge boson couplings (Fig.2). 
Here we use the following definitions for brevity,  
\bea
e^2Q^2 &\equiv& \gamma \ , \qquad
\left(g^2C_F+g^{\prime^2}Y^2-e^2Q^2\right)\equiv (W +Z) \ ,
\non\\
l &\equiv&\frac{1}{2\sqrt{2}\pi}\ln\frac{S}{\lambda^2} \ , \qquad 
L \equiv \frac{1}{2\sqrt{2}\pi}\ln\frac{S}{M^2} \ , \non
\eea
for the group factors and the loop factors respectively.
After the several cancellations among different diagrams,   
the 2 loop results for ladder diagrams, crossed ladder diagrams, 
and the diagrams including 3 point coupling turn out to be,
\bea
\mbox{ladder}&=& (1a.) + (1c.) + (1d.) + (1f.) \non\\
&=& \gamma^2 \frac{1}{24}\ l^4 +2 \gamma (W
+Z)\ \frac{1}{8}\ l^2L^2 + (W+Z)^2 \ \frac{1}{24}L^4\non\\
&& \hspace*{4cm}
+\ \underline{\gamma \ (W+Z)\left[\frac{1}{6}L^4-\frac{1}{3}L^3l
\right]} ,\non\\
\mbox{crossed} &=& (1b.) + (1e.)\times 2 +(1g.)   \non\\
&=& \gamma^2 \frac{1}{12}\ l^4 + 
\{(W+Z)^2 - g^4\frac{1}{2}C_VC_F \}\frac{1}{12}L^4\non\\   
&&\hspace{2cm} \underline{+\gamma (W+Z)\left[ -\frac{1}{6}L^4 
+\frac{1}{3}L^3 l \right]
+ 2 g^{2}e^{2}QT^{3} \left[ \frac{1}{6}L^4
 -\frac{1}{3}L^3 l \right]} ,\non\\
\mbox{3 point} &=& \left[(2a.) + (2b.) + (2c.) + (2d.)  \right]   \non\\
&=& g^4\frac{1}{2}C_VC_F \frac{1}{12}L^4
 + \underline{2 g^{2}e^{2}QT^{3} \left[ -\frac{1}{6}L^4
+\frac{1}{3}L^3 l \right]} .
\non
\eea
Here the underlined terms are the extra terms compared to QCD case 
which come from the mixing effect. After these terms cancels each
other, the sum of these contributions becomes,    
\bea
  \Gamma^{(2)} &=&  1 - 
           \frac{1}{16 \pi^2} \left( g^2 C_2 (R) + g'^2 Y^2 - e^2 Q^2 \right)
              \ln^2 \frac{S}{M^2} - \frac{1}{16 \pi^2} e^2 Q^2
                  \ln^2 \frac{S}{\lambda^2} \ , \non\\
           && + \frac{1}{2!}
            \left[\frac{1}{16 \pi^2} 
       \left( g^2 C_2 (R) + g'^2 Y^2 - e^2 Q^2 \right)
              \ln^2 \frac{S}{M^2} + \frac{1}{16 \pi^2} e^2 Q^2
                  \ln^2 \frac{S}{\lambda^2}\right]^2 \ 
\non
\eea 
which imply the exponentiation of Sudakov form factor.

We can see from this result that 
\begin{itemize}
\item[(1)]
$\gamma$ must be included to extract the correct coefficient 
of $\ln^4{\frac{S}{M^4}}$. 
\item[(2)]
Non-exponential factor appear also in ladder diagrams.  
\end{itemize}
The first point comes from the fact that the singularity from 
the photon which propagates inside the $W$ and/or $Z$ loop 
in Figs.1d and 2c is regulated by the $W$ and/or $Z$ mass \cite{M1}. 
The gauge invariant set including photon contributions must be 
included to extract the correct $\ln^2{\frac{S}{M^2}}$ terms.
The second is the important difference from QCD. In QCD case, 
the contributions of the ladder diagrams and similar part of 
the crossed ladder diagrams remain in final results and 
those of ``3-point'' diagrams act only as the counter term against 
the non-exponential term, 
while the sum of former two contributions can be extracted 
from the ordered ladder diagrams. 
Therefore, the approach by soft insetion formula which is valid for QCD 
can not be applied straightforwardly to this case and 
modifications seems to be necessary..

\begin{figure}[H]
\begin{center}
\begin{tabular}{ccc}
\leavevmode\psfig{file=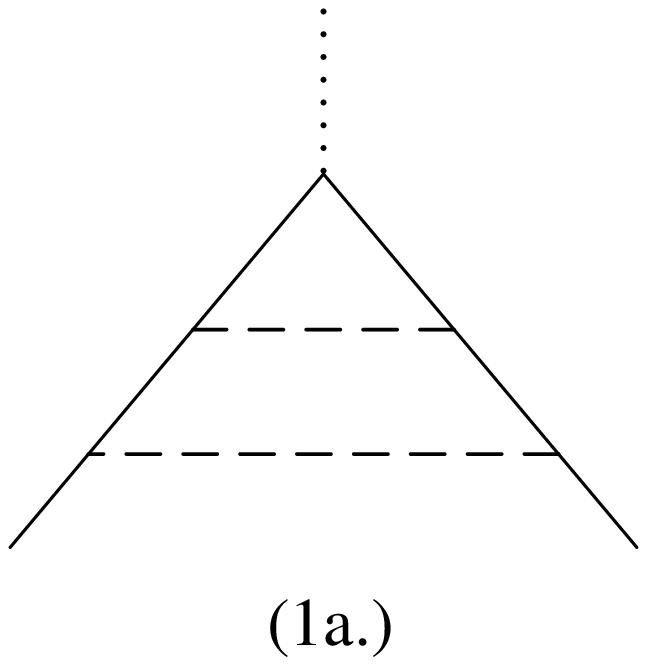,width=2cm} \quad  &
\leavevmode\psfig{file=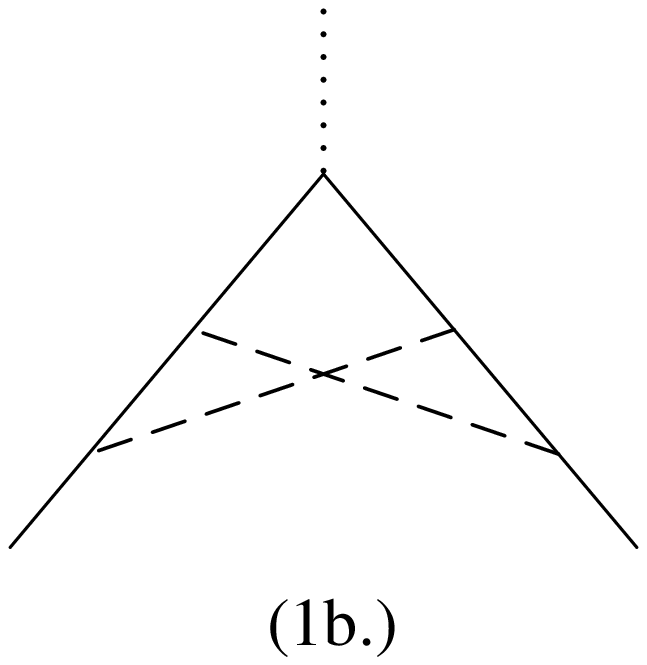,width=2cm}  & \\
& & \\
\leavevmode\psfig{file=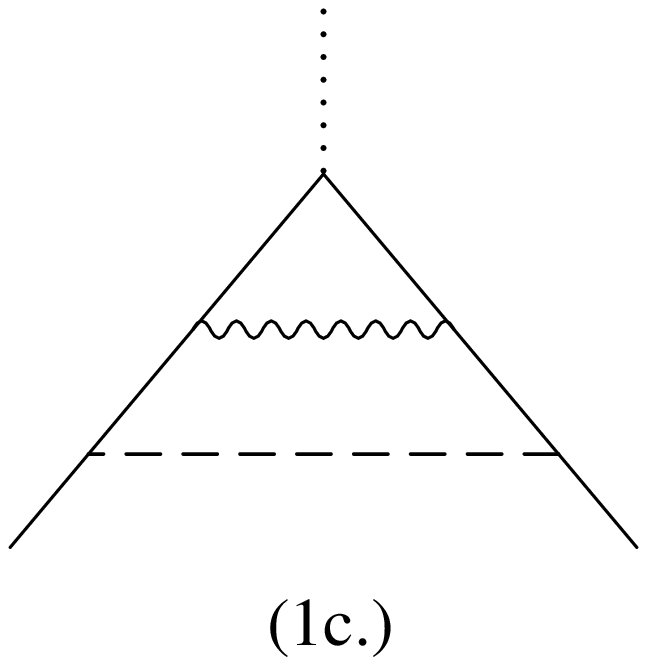,width=2cm} \quad  &
\leavevmode\psfig{file=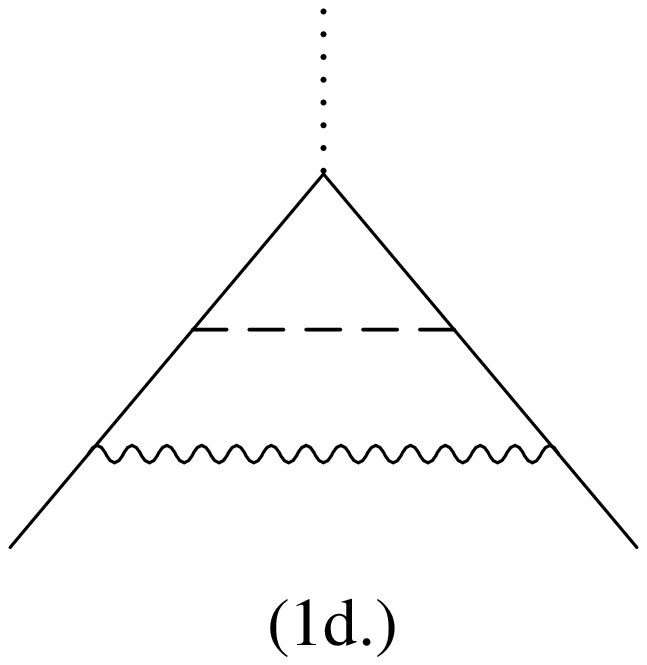,width=2cm} \quad  &
\leavevmode\psfig{file=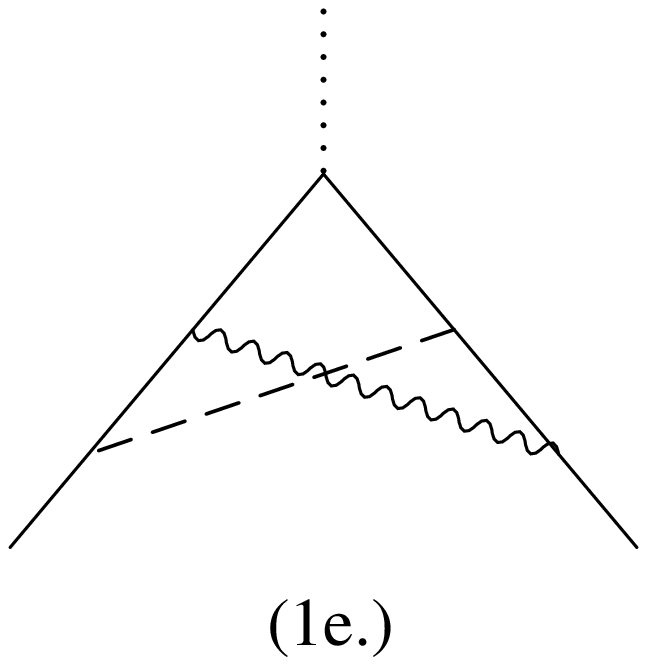,width=2cm}  \\
& & \\
\leavevmode\psfig{file=suda1f.ps,width=2cm} \quad &
\leavevmode\psfig{file=suda1g.ps,width=2cm} & 
\end{tabular}
\caption{The ladder and crossed ladder diagrams.
The dashed (wavy) line represents the photon ($W$ and/or $Z$)
with the mass $\lambda$ ($M$).}
\label{diag1}
\end{center}
\end{figure}
\vspace{-0.3cm}
\begin{figure}[H]
\begin{center}
\begin{tabular}{cccc}
\leavevmode\psfig{file=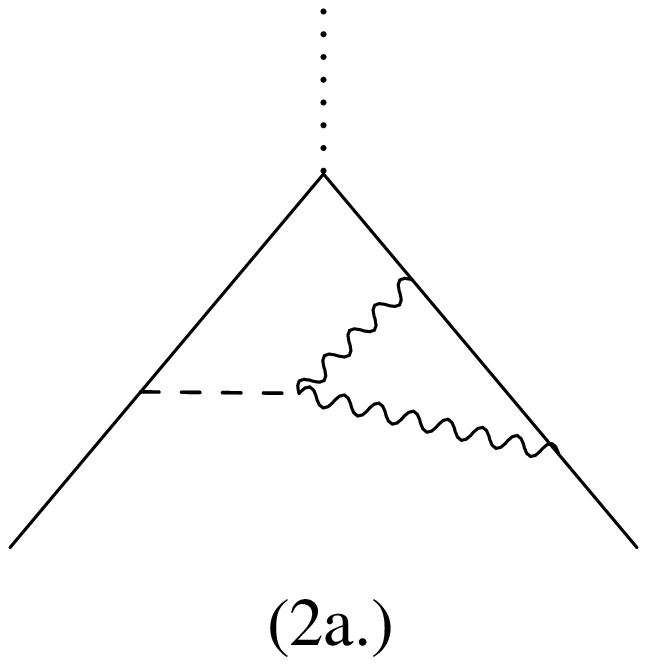,width=2cm}  &
\leavevmode\psfig{file=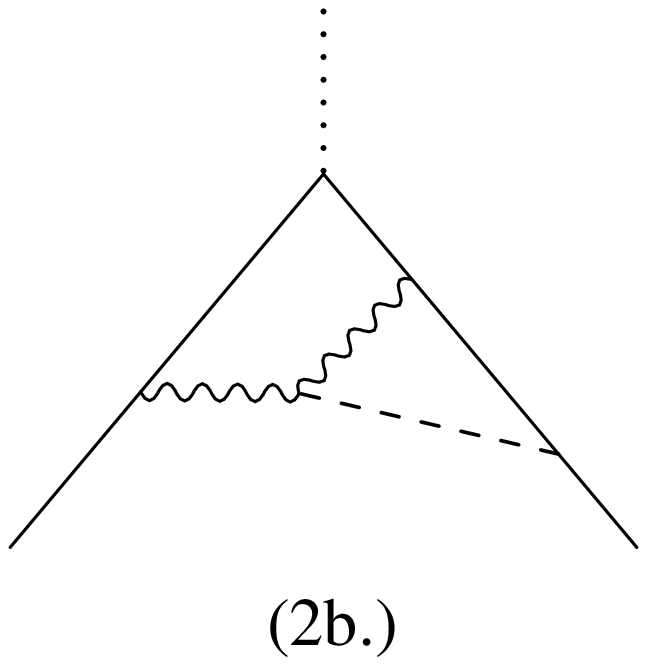,width=2cm}  &
\leavevmode\psfig{file=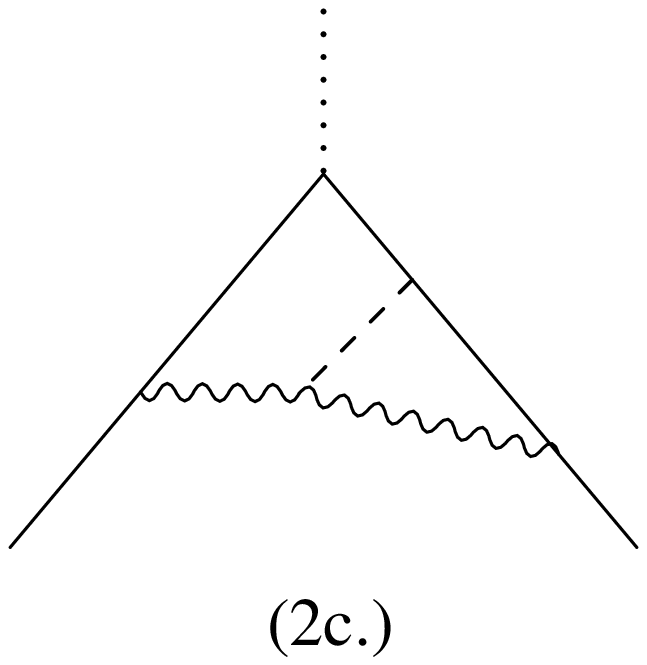,width=2cm}  &
\leavevmode\psfig{file=suda2d.ps,width=2cm}
\end{tabular}
\caption{The diagrams which have the triple point couplings.
The meaning of lines is the same as in Fig.\ref{diag1}.}
\label{diag2}
\end{center}
\end{figure}

\section{Discussions}

Electroweak log corrections become very important 
in the processes in TeV region. Since they give large corrections 
to the cross sections, It is crucial to control them in perturbation 
series to reduce the uncertainty of theoretical predictions 
in precision measurements and in the backgroud estimations for the 
signals of the new physics.\\

Exponentiation of Sudakov double logarisms seems to be valid 
from the discussion using the infrared evolution equation and 
the explicit 2-loop calculations. The resummation of next to 
leading logarisms were accomplished by K$\ddot{u}$hn, Penin and Smirnov 
\cite{KPS} for fermion external line and by Melles for general 
external line and for the top Yukawa enhanced part \cite{M2,M3}. \\

As for the phenomenology, several works have been done.
1-loop \cite{BCCVR2}, and 2-loop \cite{KPS} Sudakov-type log corrections  
to $e+e- \ra f\bar{f}$ and 1-loop effects in MSSM \cite{BRV}
were calculated. Denner and Pozzorini \cite{DP} presented the complete 
1-loop electroweak log corrections for the general processes.
Baur \cite{B} discussed on the impact of electroweak Sudakov on the W boson 
mass measurement at LHC.\\

Above calculations indicate that NLO Sudakov effects are comparable 
with LO Sudakov effects at the energy around 1 TeV. 
Therefore, it can be said that we can not yet control the electroweak 
log corrections within the 1\% level and much have to be done from now.  \\

\vspace{1cm}

\noindent
{\large \bf Acknowlegements} \\
The author would like to thank M.Hori and J.Kodaira for useful discussions 
through the collaboration. He is also grateful to H-n.Li, and 
W.M.Zhang and all organizers of PPP2000 for the kind invitation 
to a fruitful, enjoyable workshop.  The work of H.K was supported 
by the Monbusyo Grant-in-Aid for Scientific Research No.10000504.



\baselineskip 14pt            
\begin{thebibliography}{99}     
\bibitem{CC1}                           
        P. Ciafaloni and D. Comelli,                
        {\sl Phys. Lett.} {\bf B446} (1999) 278.    
\bibitem{CCC}
        M. Ciafaloni, P. Ciafaloni and D. Comelli,
        {\sl Phys. Rev. Lett.}{\bf 84} (2000) 4810; 
        {\sl Nucl. Phys.} {\bf B589} (2000) 359; 
        and hep-ph/0007096.
\bibitem{BCCRV}
        M. Beccaria, P.Ciafaloni, D. Comelli, F. Renard and C. Verzegnassi, 
        {\sl Phys. Rev.} {\bf D61} (2000) 073005. 
\bibitem{SU}                                      
        V. V. Sudakov,                              
        {\sl Sov. Phys. JETP} {\bf 3} (1956) 65.      
\bibitem{YFS}                                      
        D. Yennie, S. Frautschi and H. Suura,           
        {\sl Ann. Phys.}{\bf 13} (1961) 379.              
\bibitem{JCC}                                                       
        See {\it e.g.} J. C. Collins, in {\sl Perturbative Quantum   
        Chromodynamics} ed. A. H. Mueller (World Scientific,          
        Singapore,1989) P. 573; and references therein.              
\bibitem{S}                                                         
        See {\it e.g.} G. Sterman, in {\sl QCD and Beyond}            
        ed. D. Soper (World Scientific, Singapore, 1996)P. 327; and 
        references therein ; hep-ph/9606312.
\bibitem{CLS}
        H. Contopanagos, E. Laenen and G.Sterman, 
         {\sl Nucl. Phys.} {\bf B484} (1997) 303.
\bibitem{CC2}
        P. Ciafaloni and D. Comelli,
        {\sl Phys. Lett.} {\bf B476} (2000) 49.
\bibitem{KP}
        J. H. K\"uhn and A. A. Penin, hep-ph/9906545.
\bibitem{FLMM}
       V. S. Fadin, L. N. Lipatov, A. D. Martin and M. Melles,
       {\sl Phys. Rev.} {\bf D61} (2000) 094002.
\bibitem{KPS}
       J. H. K\"uhn, A. A. Penin and V. A. Smirnov, 
       {\sl Eur. Phys. J.} {C17} (2000) 97.
\bibitem{BW}
       W. Beenakker and A. Werthenbach, 
       {\sl Phys. Lett.}{\bf B489} (2000) 148.
\bibitem{M1}
       M. Melles, {\sl Phys. Lett.}{\bf B495} (2000) 81.
\bibitem{HKK}
       M. Hori, H. Kawamura and J. Kodaira,
       {\sl Phys. Lett.}{\bf B491} (2000) 275.
\bibitem{C}
       See {\it e.g.} M. Ciafaloni, in {\sl Perturbative Quantum
       Chromodynamics} ed. A. H. Mueller (World Scientific,
       Singapore,1989) P. 491.
\bibitem{BCM}
       A. Bassetto, M. Chiafaloni and G. Marchesini, 
       {\sl Phys. Rept} {\bf 100} (1983) 201 ; \\
       S. Catani and G.Marchesini, 
       {\sl Nucl. Phys.} {\bf B249} (1985) 301.
\bibitem{KL}
       V. N. Gribov, {\sl Yad. Fiz.} {\bf 5} (1967) 399 
       [{\sl Sov. J. Nucl. Phys.} {\bf 5} (1967) 280];\\  
       R. Kirschner and L. N. Lipatov, {\sl JETP}
       {\bf 56} (1982) 266; {\sl Phys. Rev.} {\bf D26} (1982) 1202 ;\\
       L. N. Lipatov, {\sl Nucl. Phys.} {\bf B307} (1988) 705;\\
       V. Del. Duca,  {\sl Nucl. Phys.} {\bf B345} (1990) 369.  
\bibitem{M2}
       M. Melles, {\sl Phys. Rev.} {\bf D63} (2001) 034003.
\bibitem{FS}
       R. Jakiew, {\sl Ann. Phys.} {\bf 48} 292;\\ 
       P. M. Fishbane and J. D. Sullivan, 
       {\sl Phys. Rev.} {\bf D4} (1971) 458;\\
       T. Appelquist and J. R. Primack 
       {\sl Phys.Rev.} {\bf D4} (1971) 2454  etc.
\bibitem{FFT} 
       J. Carrazone, E. Poggio and H. Quinn, 
       {\sl Phys. Rev.} {D11} 2286.   \\ 
       J. M. Cornwall and G. Tiktopoulos,
       {\sl Phys. Rev.} {\bf D13} (1976) 3370;\\ 
       J. Frenkel, M. -L. Frenkel and J. C. Taylor,
       {\sl Nucl. Phys.} {\bf B124} (1977) 268;
       and references therein. 
\bibitem{M3}
       M. Melles, hep-ph/0012157.
\bibitem{BCCRV2}
       M. Beccaria, P.Ciafaloni, D. Comelli,  F. Renard and C. Verzegnassi, 
       {\sl Phys. Rev.} {\bf D61} (2000) 011301. 
\bibitem{BRV}
       M. Beccaria, F. Renard and C. Verzegnassi, hep-ph/0007224.
\bibitem{DP}
       A. Denner and S. Pozzorini,  hep-ph/0010201.
\bibitem{B} 
       U. Baur,  hep-ph/0007287. 
%
\end{thebibliography}
\end{document}